\documentclass[10pt,conference,final]{IEEEtran}
\IEEEoverridecommandlockouts

\newcommand{\showIdentityAllowed}{}

\usepackage[T1]{fontenc}
\usepackage{cite}
\usepackage{amsmath,amssymb,amsfonts}
\usepackage{algorithmic}
\usepackage{graphicx}
\usepackage{textcomp}
\usepackage[table]{xcolor}
\usepackage[inline]{enumitem}
\usepackage{url}
\usepackage{tabularx}
\usepackage{booktabs}
\usepackage{multirow}
\usepackage{mdframed}

\newcommand{\code}[1]{\texttt{\small #1}}

\newcommand{\rqA}{What is the performance of machine learning models in predicting log placement in a large-scale enterprise system?}

\newcommand{\rqB}{What is the impact of different class balancing strategies on prediction?}
\newcommand{\rqC}{What are the most recurring relevant features across models?}

\newcommand{\rqD}{How well can a model trained with open-source data generalize to the context of a large-scale enterprise system?}

\newcommand{\apacheNum}{29}
\newcommand{\apacheJavaProdFiles}{38,980}
\newcommand{\apacheLogStatements}{55,849}
\newcommand{\apacheMethods}{388,086}
\newcommand{\apacheLoggedMethods}{27,247}

\newcommand{\adyenJavaProdFiles}{34,526}
\newcommand{\adyenLogStatements}{50,981}
\newcommand{\adyenMethods}{309,527}
\newcommand{\adyenLoggedMethods}{23,723}
\newcommand{\adyenLogRatio}{7.7}

\ifdefined\showIdentityAllowed
\newcommand{\adyen}{Adyen}
\else
\newcommand{\adyen}{COMPANY}
\fi

\begin{document}

\title{An Exploratory Study of Log Placement Recommendation in an Enterprise System}

\ifdefined\showIdentityAllowed

\author{
    \IEEEauthorblockN{
        Jeanderson C\^andido\IEEEauthorrefmark{1}\IEEEauthorrefmark{2},
        Jan Haesen\IEEEauthorrefmark{2},
        Maur\'icio Aniche\IEEEauthorrefmark{1}, and
        Arie van Deursen\IEEEauthorrefmark{1}
    }
    \IEEEauthorblockA{
        \IEEEauthorrefmark{1}Department of Software Technology\\ Delft University of Technology, The Netherlands\\
        \{j.candido, m.f.aniche, arie.vandeursen\}@tudelft.nl
    }
    \IEEEauthorblockA{
        \IEEEauthorrefmark{2}Adyen N.V., The Netherlands\\
        \{jeanderson.candido, jan.haesen\}@adyen.com
    }
}

\else

    \author{\IEEEauthorblockN{[Anonymous authors]}}

\fi

\maketitle

\begin{abstract}

Logging is a development practice that plays an important role in the operations and monitoring of complex systems.
Developers place log statements in the source code and use log data to understand how the system behaves in production.
Unfortunately, anticipating where to log during development is challenging.
Previous studies show the feasibility of leveraging machine learning to recommend log placement despite the data imbalance since logging is a fraction of the overall code base.
However, it remains unknown how those techniques apply to an industry setting, and little is known about the effect of imbalanced data and sampling techniques.
In this paper, we study the log placement problem in the code base of \adyen{}, a large-scale payment company.
We analyze \adyenJavaProdFiles{} Java files and \adyenMethods{} methods that sum up +2M SLOC.
We systematically measure the effectiveness of five models based on code metrics, explore the effect of sampling techniques, understand which features models consider to be relevant for the prediction, and evaluate whether we can exploit \apacheMethods{} methods from \apacheNum{} Apache projects to learn where to log in an industry setting. 
Our best performing model achieves 79\% of balanced accuracy, 81\% of precision, 60\% of recall.
While sampling techniques improve recall, they penalize precision at a prohibitive cost.
Experiments with open-source data yield under-performing models over \adyen{}’s test set; nevertheless, they are useful due to their low rate of false positives.
Our supporting scripts and tools are available to the community.
\end{abstract}


\section{Introduction}

Logging is a popular development practice that plays an important role in the operations and monitoring of complex systems.
Developers place log statements in the source code to expose the internal state of the system as it runs in production.
Usually, companies rely on a logging infrastructure (e.g., Elastic Stack~\cite{elastice_elastic_nodate}) to collect and process that data either in-house or as a cloud service.
Operations and monitoring rely on metrics and data collected at different layers of the stack, from hardware, operating system, and web server logs to the application logs themselves.
That data is necessary for several log analysis tasks, e.g., anomaly detection~\cite{meng_loganomaly_2019,zhang_robust_2019, du_deeplog_2017}, root cause analysis~\cite{lu_log-based_2017,gurumdimma_crude_2016}, and performance monitoring~\cite{yao_log4perf_2018}.

In practice, developers still rely on their intuition to determine which parts of the system require log statements~\cite{li_qualitative_2020}.
This problem is known as the \emph{``where to log''} (or \emph{``log placement''}) problem~\cite{fu_where_2014,zhu_learning_2015,li_where_2020,li_studying_2018}.
Empirical studies show that developers continuously spend time fixing and adjusting log statements in a trial-and-error manner~\cite{yuan_characterizing_2012,chen_characterizing_2017}.
In addition, improper logging also leads to several issues in production software and logging infrastructure~\cite{hassani_studying_2018}.
Deliberately placing log statements everywhere in the source code might increase data throughput, resulting in more demand for storage and processing time for data indexing.
Conversely, missing log statements undermine the ability to diagnose failures and abnormal behavior.
In an enterprise context, the issues associated with improper logging might not only reflect in costs with commodity or on-premise resources but also in the company's reputation since failing to provide timely response to failures and abnormalities can be harmful to the business.

The research community has been proposing techniques to support developers in deciding what parts of the system to log.
For instance,
Jia~et~al.~\cite{jia_smartlog_2018} proposed an approach based on association rule mining to place error logs on \code{if} statements;
Li~et~al.~\cite{li_studying_2018} studied the use of topic modeling for log placement at method-level, and;
Li~et~al.~\cite{li_where_2020} proposed a deep learning-based approach to indicate the need for logging at block-level.
Those techniques rely mostly on code vocabulary to learn placement patterns in the source code, and experiments with open-source data show their relevance on the placement problem.
Unfortunately, it remains unknown how existing approaches apply in an enterprise setting.
Zhu~et~al.~\cite{zhu_learning_2015} proposed a technique and evaluated it on two large-scale systems; however, it operates on specific contexts:  \code{catch} clauses (exception handling) and \code{if} statements with return values.
In practice, there are many different types of logging (e.g., exceptional and unexceptional cases)~\cite{fu_where_2014} and, in a business-critical application, every log is under scrutiny for monitoring and postmortem analysis.
Furthermore, existing techniques employ over-sampling techniques (e.g.,~\cite{zhu_learning_2015,li_where_2020}) to avoid bias towards the majority class; however, it remains unknown the effect of imbalanced data and sampling techniques (i.e., over-sampling of the minority class or under-sampling of the majority class) on log placement.

In this paper, we study the log placement problem in the code base of \adyen{}, a large-scale payment company.
We analyze \adyenJavaProdFiles{} Java files and \adyenMethods{} methods that sum up to over two million lines of code.
We address log placement as a supervised binary classification problem at method-level.
We focus at method-level given the low prevalence of log statements on logged methods (up to 75\% of logged methods contain no more than two log statements) and the diverse contexts of log placement (more details in Section~\ref{sec:motivation}).
Our intuition is that we can trade the specificity of block-level prediction without losing awareness.
We leverage object-oriented metrics and structural metrics as predictors to placement patterns since they are easy to compute via static analysis. In addition, they have been widely used in other domains (e.g., defect prediction~\cite{trautsch_static_2020,zimmermann_cross-project_2009,turhan_relative_2009,dambros_extensive_2010,taba_predicting_2013}) but are under explored for log placement.
We systematically measure the effectiveness of five machine learning models, explore the effect of sampling techniques, understand which features models consider to be relevant for the prediction, and evaluate whether we can leverage code metrics and open-source data to learn log placement in an industry setting.

Our results show that
\begin{enumerate*}[label=(\roman*), font=\itshape]
\item our best performing model (Random Forest) achieves $79\%$ of balanced accuracy, $81\%$ of precision, $60\%$ of recall, and $1\%$ of false positive rate,
\item sampling techniques improve recall ($+26\%$ up to $+52\%$), but at a prohibitive cost in precision ($-18\%$ up to $-40\%$),
\item the depth of a method, i.e., maximum number of nested blocks, is a key indicator of log placement, and
\item models based on open-source data yield lower performance on \adyen{}'s test set; nevertheless, they can be used to overcome the cold-start problem due to their low rate of false positives
\end{enumerate*}.
This paper provides the following contributions:
\begin{itemize}
\item A feature engineering process for machine learning based on code metrics for log placement at method-level;
\item An empirical evaluation of five machine learning models and their performance on a large-scale enterprise system;
\item An analysis of the effectiveness of transfer learning from open-source data to overcome the cold-start problem in an industry setting; 
\item A fully functional toolkit for auditing and extension~\cite{candido_electronic_2021}.
\end{itemize}

\section{Context and Motivation}
\label{sec:motivation}

\adyen{} offers payment processing as a service and connects shoppers to more than 4.5k merchants at global scale.
At the core of the business, there is a platform with more than 10 years of development that is maintained and evolved by hundreds of developers around the globe.
Technology plays a key role to scale hundreds of transactions per second in different currencies and different payment channels (e.g., points of sale terminals and e-commerce).
\ifdefined\showIdentityAllowed
Only in 2019, \adyen{} processed 240 billion euros in transaction volume\footnote{\url{https://www.adyen.com/investor-relations/H2-2019}~(Accessed on September 2020).}.
\fi

To ensure quality in a rapid growing business, developers follow DevOps practices and are responsible for testing and monitoring their changes on every release.
Moreover, there are dedicated infrastructure and monitoring teams responsible for, among other activities, building and maintaining data pipelines to support system monitoring.
In a monthly basis, the in-house clusters process billions of log events generated by the platform.
The use of this volume of data goes beyond monitoring the stability of release changes.
For instance, teams leverage that data to trace fraudulent activity, diagnose abnormal behavior, and perform root cause analysis.
Good logging practices and efficient monitoring are vital to the business operation and scale of \adyen.

In the following, we describe the pervasiveness of log statements in the code base (Section~\ref{sec:motivation:subject-and-logstatements}), how developers conduct log engineering (Section~\ref{sec:motivation:practices}), and the motivation for an automated approach (Section~\ref{sec:motivation:mlse}).

\subsection{Pervasiveness of Log Statements}
\label{sec:motivation:subject-and-logstatements}

\begin{table}[!t]
\caption{Placement of log statements grouped by direct enclosing context.}
\footnotesize
\begin{center}
\begin{tabularx}{\columnwidth}{@{}Xrr@{}}
\toprule
Enclosing context & Log statements &  \% \\
\midrule

\code{if-else} statements & 27,257 &  53 \\
     \code{catch} clauses & 14,935 &  29 \\
      method declarations &  5,823 &  11 \\
    \code{try} statements &  1,464 &   3 \\
          loop statements &    935 &   2 \\
                   others &    567 &   1 \\

\midrule
Total & \adyenLogStatements & 100 \\
\bottomrule
\end{tabularx}
\label{tab:log-placement}
\end{center}
\end{table}

\begin{figure}[!t]
\centerline{\includegraphics[width=\columnwidth]{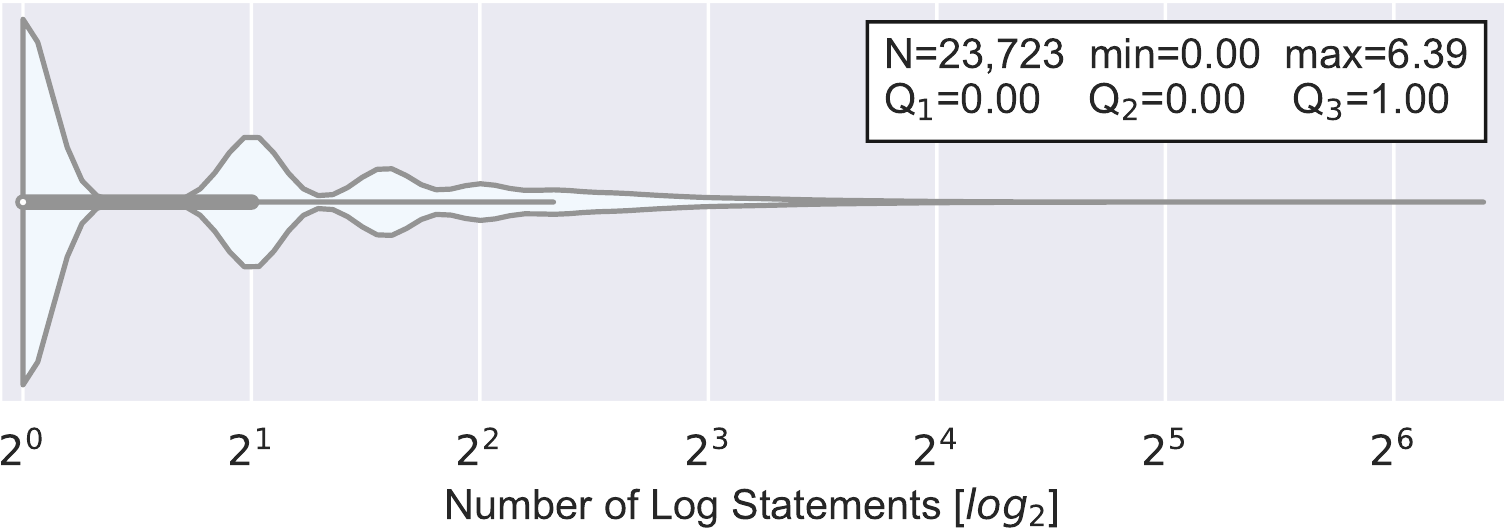}}
\caption{Distribution of log statements per logged method (in $log_2$ scale). Each data point represents the number of log statements in a method. 75\% of the logged methods contain up to two log statements.}
\label{fig:dist-num-logstmts}
\end{figure}

To analyze the presence of log statements,
we focused on production-related code due to the importance of log data in the field
(we elaborate more details about measurements in Section~\ref{sec:methodology:subjects}).
Our dataset contains \adyenJavaProdFiles{} Java files and \adyenMethods{} methods that sum up to more than two million lines of source code.
Only \adyenLogRatio{}\% of the methods (\adyenLoggedMethods{} out of \adyenMethods{}) contain at least one log statement.
Figure~\ref{fig:dist-num-logstmts} shows the distribution of log statements among those methods;
we used log scale to highlight the distribution modes.
In most cases, the number of log statements per logged method is relatively low: in 75\% of the cases, there are up to two log statements and 50\% of the methods contain only one log statement.
To further understand the placement context of log statements, we also analyzed the abstract syntax tree from those methods.
The most common placements are \code{if-else} statements, followed by \code{catch} clauses, and method declarations (Table~\ref{tab:log-placement}).

Overall, the percentage of production-related methods that contain log statements is relatively low (7.7\%).
While the frequency of log statements in methods is low (up to two in 75\% of the cases), log statements occur in many different contexts in a method body (Table~\ref{tab:log-placement}).

\subsection{Log Engineering in the Field}
\label{sec:motivation:practices}
\adyen{} has internal guidelines to educate developers about logging practices (e.g., log levels and message formats) and how to use the in-house logging framework.
These guidelines help to align the perception about logs for developers working on different teams and with different backgrounds.
A major challenge in this context is to add log statements with enough contextual information without overloading the data infrastructure.

To better understand the maintenance effort on log statements, we conducted an exploratory analysis of log-related changes in the repository from November 1\textsuperscript{st}, 2019 to April 30\textsuperscript{th}, 2020.
We used a keyword-based approach to identify log-related commits and leveraged the structured titles from commits to identify associated tasks from the issue tracker for additional context.
Commit titles are in the format ``$<$Task-id$>$ $<$Title$>$'', where ``$<$Task-id$>$'' is a mandatory field.
We report relative values to protect confidential company data.
In the following we summarize our observations:

\begin{itemize}
\item \textbf{Log-related commits are recurrent:}
We grouped the log-related commits per month and computed their relative frequency to Java-related commits.
On average, there were $7.0\pm1.6\%$ of commits focused exclusively on log-related changes (min=$5.7\%$ and max=$9.7\%$).
This suggests that there is a recurrent demand to improve log-related code.

\item \textbf{Most log-related commits are isolated small fixes:}
We analyzed the commits identified previously to understand what was the context of the change.
We used the ``$<$Task-id$>$'' field to identify whether the change was associated to an open task or an isolated fix (tagged as ``FIX'').
The percentage of isolated fixes was almost twice ($\times1.94$) as high compared to task-related commits.
Previous studies have also observed the same phenomenon in open-source projects~\cite{chen_characterizing_2017,yuan_characterizing_2012}.
In an ideal setting, the developer should be able to anticipate the need of those small and isolated fixes to improve productivity.

\end{itemize}

\subsection{The Need For an Automated Approach}
\label{sec:motivation:mlse}

Our observations about log engineering in the field suggest that, while there is a company-wide effort to strike a balance between high-quality data and cost-effective usage of infrastructure,
there is a need for tooling support to make consistent logging decisions.
Unfortunately, while experts might provide assistance during code-review, this approach does not scale.
Furthermore, it is unfeasible for a developer to have a contextual baseline of comparison between an ongoing task and the hidden patterns of log statements spread in a large code base.

As a first step towards alleviating the burden of implementing and maintaining logging code at \adyen{}, in this work, we explore the feasibility of exploiting machine learning on the complex task of log placement.
Our intuition is that non-trivial methods are more likely to be logged but not necessarily all complex methods should be logged.
Ideally, \emph{if the developer introduces a method similar in complexity to existing methods that contain log statements, then, the developer should be aware of this relationship.}
We believe this awareness might help the developer to make informed decisions before the code deploys to production.

We define the scope of prediction at method level since we observed that, in our context, log statements are not recurrent in logged methods, e.g. 50\% of the logged methods contain only one log statement  (Figure~\ref{fig:dist-num-logstmts}).
In addition, as seen previously, log statements are spread in many different contexts (Table~\ref{tab:log-placement}).

\section{Methodology}
\label{sec:methodology}

Our goal is to predict whether a method should be logged or not in an enterprise system.
To this end, we frame this problem as a supervised binary classification problem at method level.
We analyze the feasibility of using code metrics as predictors, evaluate the performance of different machine learning models, the relevance of class balancing, and whether models built on top of open-source data generalise to our domain.
Concretely, we investigate the following research questions:

\begin{enumerate}[label=\textbf{\footnotesize RQ\arabic*:}, align=left, left=0pt, itemsep=4pt, topsep=4pt]
\item \textbf{\rqA{}} This $RQ$ addresses the feasibility of using code metrics as predictors.
Our initial assumption is that, given a sufficient number of training examples, a model can differentiate a method that should be logged or not based on the distribution of features extracted from methods and their enclosing class. Answering this $RQ$ is important to validate this assumption.

\item \textbf{\rqB{}} This $RQ$ addresses the problem of imbalanced data in classification problems~\cite{he_learning_2009,bennin_investigating_2016}.
Imbalanced data occurs when the frequency of one class is significantly smaller than the other(s).
In our context, only \adyenLogRatio{}\% of methods in the code base have log statements (see Section~\ref{sec:motivation:subject-and-logstatements}).
This might introduce a bias towards the majority class and result in an underperforming model.
Answering this $RQ$ is important to assess the impact of imbalanced data in our domain and the trade-offs of different sampling techniques.

\item \textbf{\rqC{}} This $RQ$ addresses the relevance of features in the task of differentiating methods that have log statements from methods without log statements.
In our experiments, we use machine learning algorithms that learn from data on different ways.
Our goal is to understand if those differences cause the models to capture different characteristics from data.
Answering this research question is important to provide insights about what models learn after training.

\item \textbf{\rqD{}} 
This $RQ$ addresses the feasibility of learning from open-source data.
Transfer learning has been applied in other domains (e.g., defect prediction~\cite{zimmermann_cross-project_2009,nam_transfer_2013,zhang_empirical_2015,tong_kernel_2019} and performance modeling~\cite{jamshidi_transfer_2017, moradi_performance_2019,ha_deepperf_2019}) to overcome the ``cold-start problem'', i.e., when it is not possible to train a model due to insufficient (or nonexistent) training data.
In our context, this relates to a project being in an early stage of development, for instance.
Answering this $RQ$ is important to understand transfer learning for log placement.

\end{enumerate}

\begin{table}[bp]
\caption{Summary of data collection.}
\footnotesize
\begin{center}
\begin{tabularx}{\linewidth}{@{}Xrrrrr@{}}
\toprule
Dataset
    & Files$^a$
    & \multicolumn{1}{>{\raggedleft}m{12mm}}{Log statements}
    & \multicolumn{1}{>{\raggedleft}m{10mm}}{Methods (\textsc{\tiny M})}
    & \multicolumn{1}{>{\raggedleft}m{12mm}}{Logged methods (\textsc{\tiny LM})}
    & $\frac{\textrm{\textsc{LM}}}{\textrm{\textsc{M}}}\textrm{(\%)}$
    \\

\midrule
\adyen & \adyenJavaProdFiles & \adyenLogStatements & \adyenMethods & \adyenLoggedMethods & \adyenLogRatio \\
Apache ({\tiny N=\apacheNum}) & \apacheJavaProdFiles & \apacheLogStatements & \apacheMethods & \apacheLoggedMethods & -- \\
\midrule
\multicolumn{3}{@{}l}{Apache summary:}\\
\multicolumn{1}{r}{Min}    & 117    & 120     & 1,176   & 54     & 4.2 \\
\multicolumn{1}{r}{Q1}     & 586    & 727     & 4,839   & 417    & 4.9 \\
\multicolumn{1}{r}{Median} & 821    & 1,514   & 6,933   & 762    & 7.1 \\
\multicolumn{1}{r}{Average}& 1,344  & 1,926   & 13,383  & 940    & 8.1 \\
\multicolumn{1}{r}{Q3}     & 1,577  & 2,345   & 14,915  & 1,154  & 9.9 \\
\multicolumn{1}{r}{Max}    & 5,881  & 12,147  & 65,184  & 4,955  & 15.8\\
\bottomrule
\multicolumn{6}{@{}l}{$^a$Production-related Java code.}
\end{tabularx}
\label{tab:subjects:summary}
\end{center}
\end{table}

\textbf{Tooling Support:}
All supporting scripts and tools are available in our online appendix~\cite{candido_electronic_2021}.
The \code{README.md} file located in the root directory describes the components  and the minimum requirements.

In the following, we describe the data collection, labelling, feature extraction, learning algorithms, machine learning process, and how we address each research question.

\subsection{Data Collection}
\label{sec:methodology:subjects}

The focus of our study is the code base of our industry partner.
In addition, we create a dataset based on \apacheNum{} projects from the Apache ecosystem to address the feasibility of learning where to log from them.
We focused on Apache projects because they are often present in industry-level projects at different domains (e.g., cloud managing, databases, web servers, and big data solutions).
Our assumption is that they might contain good or influential logging practices since their users rely on log data for troubleshooting and monitoring the execution of complex systems.
Table~\ref{tab:subjects:summary} summarizes the datasets and provides an overview of the pervasiveness of log statements.
In the following we describe how we analyze the subjects and the selection criteria to compose our dataset.

\textbf{Source code filtering:} For all projects, including \adyen's code base, we selected Java files that are related to production code.
The rationale for excluding non-production sources is that those files contain methods that may not reflect the developer's perception of production level logging decisions.
We searched for files with the suffix \code{.java} using the \code{find} UNIX utility~\cite{freebsd_find1_nodate} and grouped the entries into the following categories: \begin{enumerate*}[label=\emph{(\roman*)}] \item test-related files, \item documentation-related files, e.g., ``how to'' recipes and API examples, and \item build-related files \end{enumerate*}.
We classify each Java file by checking whether its corresponding path
contains a keyword related to one of the categories.
For test-related files, the keywords are ``\emph{fixture}'', ``\emph{memtest}'',
``\emph{/mock/}'', and ``\emph{test/}''.
For documentation-related files, the keywords are ``\emph{docs/}'', ``\emph{/examples/}''.
For build-related files, the keyword is ``\emph{buildSrc/}'' which corresponds to the path of custom Gradle tasks.
We classify the remaining files as production-related file (Table~\ref{tab:subjects:summary}, column ``Files'').

\textbf{Log statements:} In the subset of production-related files, we measure the pervasiveness of log statements with static analysis.
For each method in the source code, we visit all method invocations in the enclosing method body and check whether it represents an expression statement. This allows us to identify multi-line log statements and log statements that use fluent API patterns (a.k.a., method chaining) which are supported by popular logging frameworks (e.g., Slf4j~\cite{slf4jorg_slf4j_nodate} and Log4j~2~\cite{apache_log4j_log4j_nodate}).
We use the lowercase string representation of the expression to check if the expression matches the patterns of popular log APIs (i.e., ``\code{.*log(ger)?.*}'' or ``\code{.*(error|warn|info|debug).*}'').
In case it matches, we increment the counter of log statements and mark the current method as a logged method.
Columns ``{\footnotesize Log statements}'', ``{\footnotesize Methods (\textsc{\tiny M})}'', ``{\footnotesize Logged methods (\textsc{\tiny LM})}'', and ``{\footnotesize $\frac{\textrm{LM}}{\textrm{M}}\textrm{(\%)}$}'' summarize the counting of log statements, visited methods, logged methods, and percentage of logged methods, respectively.

\textbf{Apache selection:} We selected Java projects listed in the official website from Apache foundation\footnote{\url{https://projects.apache.org/projects.html?language}}.
We initially downloaded 64 projects that we were able to clone automatically based on the scrapped data from their website.
We excluded projects with less than or equals to 4\% of logged methods and projects with less than or equals to 100 production-related files.
The rationale for the log percentage criterion is to define a lower bound cutoff for projects with less than half of the average percentage of logged methods from \adyen{} dataset ($\adyenLogRatio{}\%\times\frac{1}{2}\approx4\%$); the rationale for the number of production-related files criterion is to exclude small library projects.
After the selection criteria, we reduced the original list of 64 to \apacheNum{} projects in the final set\footnote{Selected projects available at ``\texttt{\url{./apache-projects-paper.csv}}''~\cite{candido_electronic_2021}}.

The selected Apache projects sum up to \apacheJavaProdFiles{} Java files with an average of 8\% of logged methods (see ``Apache summary'' in Table~\ref{tab:subjects:summary} for the detailed distribution).
The project with the highest percentage of logged methods is Apache Sqoop, a system designed for data transmission which has a percentage of 15.8\% of logged methods (488 out of 3,080) and 351 production-related files.
The project with the lowest percentage of logged method is Apache Ignite, an in-memory database and caching platform which has a percentage of 4.2\% of logged methods (2,765 out of 65,184) and 5,881 Java files.

\subsection{Label Identification and Feature Extraction}
\label{sec:methodology:fe-and-labelling}

\begin{figure*}[hbtp]
\centerline{\includegraphics[width=\linewidth]{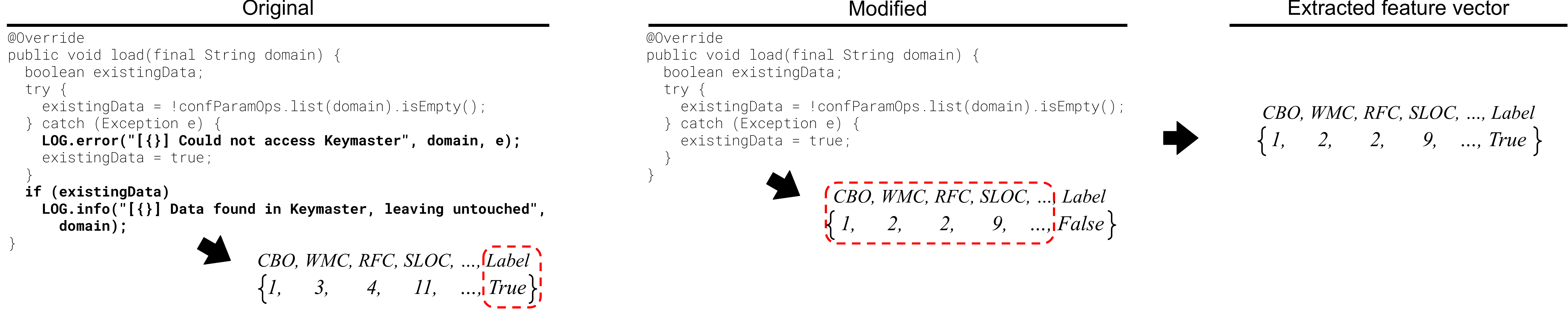}}
\caption{Label identification and feature extraction process. The ``Extracted feature vector'' for the corresponding method consists of the label identified in the original code snippet and the features calculated after removing log statements (i.e., ``Modified'' code snippet). Class features omitted for brevity.}
\label{fig:feature-eng}
\end{figure*}

We annotate the labels based on the presence of log statements in the methods identified during data collection, and we use code metrics as features to characterize methods in the dataset.
Code metrics have been widely used in other domains (e.g., defect prediction~\cite{trautsch_static_2020,     zimmermann_cross-project_2009,turhan_relative_2009,dambros_extensive_2010,taba_predicting_2013}) and are simple to compute using static analysis.
Furthermore, the use of code metrics allows us to start with a simple feature extraction pipeline as it is unnecessary to employ tokenization, word embedding, and other techniques from Natural Language Processing (NLP).

\textbf{Features:} Table~\ref{tab:metrics} highlights the 45 code metrics we use as features: 40 structural metrics (38 quantitative and two qualitative) and five object-oriented metrics~\cite{chidamber_towards_1991,chidamber_metrics_1994}.
The 38 quantitative structural metrics indicate the frequency of AST nodes (e.g., number of parameter declarations in a method), except lines of source code (``SLOC''), and the two qualitative metrics indicate the type of a compilation unit (``classType'') and whether a method is a constructor (``isConstructor'').
Compilation units are classified either as ``enum type'', ``class'', ``inner class'', ``interface'', or ``anonymous''.
We reuse an open-source tool available on GitHub based on Eclipse JDT~\cite{aniche_java_2015} to calculate the metrics.

\begin{table}[!b]
\caption{Summary of code metrics used as features.}
\footnotesize
\begin{center}
\begin{tabularx}{\columnwidth}{X@{}}
\toprule
\multicolumn{1}{c}{Code metrics\textsuperscript{a}}
\\
\midrule
\multicolumn{1}{@{}l}{Object-oriented metrics ($\textrm{N}=5$):}
\\
{
\textsc{CBO}, \textsc{DIT}, \textsc{WMC}, \textsc{RFC}, \textsc{LCOM}
}
\\
\multicolumn{1}{@{}l}{Qualitative ($\textrm{N}=2$):}
\\
{
{isConstructor}, {classType}
}
\\
\multicolumn{1}{@{}l}{Quantitative ($\textrm{N}=38$):}
\\
\multicolumn{1}{X}
{
{abstractMethodsQty},
{anonymousClassesQty},
{assignmentsQty},
{comparisonsQty},
{defaultFieldsQty},
{defaultMethodsQty},
{finalFieldsQty},
{finalMethodsQty},
{innerClassesQty},
{lambdasQty},
{SLOC},
{loopQty},
{mathOperationsQty},
{maxNestedBlocksQty},
{methodsInvokedIndirectLocalQty},
{methodsInvokedLocalQty},
{methodsInvokedQty},
{NOSI} (number of static invocation),
{numbersQty},
{parametersQty},
{parenthesizedExpsQty},
{privateFieldsQty},
{privateMethodsQty},
{protectedFieldsQty},
{protectedMethodsQty},
{publicFieldsQty},
{publicMethodsQty},
{returnsQty},
{staticFieldsQty},
{staticMethodsQty},
{stringLiteralsQty},
{synchronizedFieldsQty},
{synchronizedMethodsQty},
{totalFieldsQty},
{totalMethodsQty},
{uniqueWordsQty},
{variablesQty},
{visibleFieldsQty}
}
\\
\bottomrule
\multicolumn{1}{@{}l}{\textsuperscript{a}Check the tool's GitHub page for further details~\cite{aniche_java_2015}.}
\end{tabularx}
\label{tab:metrics}
\end{center}
\end{table}

\textbf{Feature extraction:} We combine method-level metrics with metrics from their enclosing class.
Our intuition is that this might help to differentiate data points and provide further context to learning algorithms.
There are 63 features in total after extracting the 45 metrics at method and class-level.
Note that some metrics are specific to classes (e.g., type of a compilation unit), some metrics are specific to methods (e.g., whether a method is a constructor), and others apply to both levels (e.g., \textsc{SLOC} from a given method and \textsc{SLOC} from the enclosing class).
The final dimensionality is 68 features after the encoding of the categorical variables: 61 numerical features, two features that indicate whether a method is a constructor, and five features that indicate the compilation type.

\begin{figure*}[!ht]
\centering
\includegraphics[width=\linewidth]{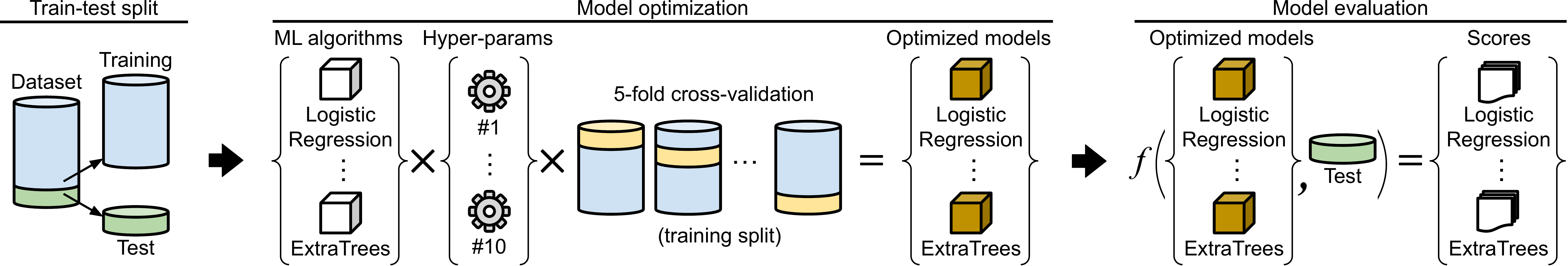}
\caption{Machine learning process: for a given learning algorithm, we use 10 random selections of hyper-parameters (``Hyper-params'') and five-fold cross-validation for model optimization; we compute the performance metrics (``Scores'') using an independent test set to compare the optimized models.}
\label{fig:ml-pipeline}
\end{figure*}

It is worthy mentioning that feature extraction requires code transformation to avoid \emph{data leakage} caused by the existing log statements.
Log statements might impact the distribution of the collected metrics due to their implicit dependency with some features, e.g., lines of source code (``SLOC'') and number of method invocations (``methodsInvokeQty'').
Therefore, we remove log statements and guards that might exist whenever possible\footnote{Details available at ``\texttt{\url{./log-remover}}''~\cite{candido_electronic_2021}}.

In addition, the presence of \code{try-catch} blocks is another potential source of data leakage.
Recall from Table~\ref{tab:log-placement} that 32\% of log statements at \adyen{}'s dataset are direct children of \code{try-catch} statements.
Data leakage caused by \code{try-catch} statements occurs when the developer handles an unchecked exception to add a log statement in the catch clause and re-throws the exception.
While our intuition is that those are rare cases, we make the conservative decision of not using the presence (or frequency) of \code{try-catch} as a feature. 
The output of feature extraction is a matrix: each row corresponds to a feature vector that represents a given method and each column corresponds to a feature.

We summarize this step using a simplified version of the method ``load'' from Apache Syncope, an identity manager for enterprise systems (Figure~\ref{fig:feature-eng}).
The highlighted lines in the original code snippet indicate the presence of log statements, therefore, ``\emph{Label}'' is ``\emph{True}''.
Next, we remove log statements (``Modified'' code snippet) and extract the code metrics (class metrics are omitted for brevity).
The resulting feature vector for ``load'' is the combination of code metrics without log statements and the identified label.
We index labels and features by file path, the fully-qualified class name, and method signature to avoid misplacing labels.
Figure~\ref{fig:feature-eng} also highlights the difference of feature vectors caused by the presence of the log statements in the original code snippet.

\subsection{Learning Algorithms and Machine Learning Process}
\label{sec:methodology:ml}

In our experiments, we use five classifiers:
\begin{enumerate*}[label=(\roman*), font=\textit]
\item Logistic regression~\cite{bishop_pattern_2006} is a simple classifier based on linear regression and the sigmoid function;
\item Decision Tree~\cite{quinlan_c45_1993} derives a criteria that best split the data among the output classes;
\item Random Forest~\cite{breiman_random_2001}, \item AdaBoost~\cite{hastie_multi-class_2009}, and \item Extremely Randomized Trees (Extra Trees)~\cite{geurts_extremely_2006} are tree-based ensemble approaches that combine a set of Decision Trees to make predictions.
\end{enumerate*}
It is worthy mentioning that the tree-based models are insensitive to \emph{feature scaling} and implicitly performs \emph{feature selection}.
For logistic regression, we handle scaling as a hyper-parameter with five options\footnote{The details about the scalers and their effect are available in the \code{\footnotesize scikit-learn} documentation~\cite{python_compare_nodate}}: \begin{enumerate*}[label=(\roman*), font=\textit]\item {MinMaxScaler}, \item {Normalizer},  \item {RobustScaler}, and \item {StandardScaler}\end{enumerate*}.
In addition, we use logistic regression with ridge regularization (L2) by default.
We refer the reader to the papers cited above for a better understanding of each algorithm.
All algorithms are available in the \code{scikit-learn} library~\cite{pedregosa_scikit-learn_2011}.

Figure~\ref{fig:ml-pipeline} illustrates our machine learning process.

\textbf{Train-test split:} For a given dataset, we shuffle and split it into training (80\%) and test (20\%) sets. The split is stratified, i.e., it preserves the original class balance. The rationale for this step is twofold: first, it allows us to find the best hyper-parameters and evaluate the optimized model on independent sets from the same data distribution; second, it allows us to reuse a test set to compare the performance of models trained on different datasets.

\textbf{Model optimization:} 
For each learning algorithm, we try different hyper-parameters and select the values that yield the best performance on five-fold cross-validation.
We optimize for balanced accuracy ($BA$) as the data is imbalanced (more details in ``Model Evaluation'').
The hyper-parameter values are randomly selected according to set of possible values defined \emph{apriori}.\footnote{Detailed settings available at ``\texttt{\url{./log-prediction/logpred_method/models.py}}''~\cite{candido_electronic_2021}}
For instance, if using AdaBoost, the possible values for \emph{learning rate} are $10^{\{0, -1, -2, -3, -4\}}$.
The rationale for random selection is that exhaustively trying all possible values for all possible hyper-parameters is costly due to cross-validation; hence, we limit the trials to 10 runs.

\textbf{Model evaluation:}
For each optimized model, we compute the scores of the following performance metrics:
\begin{itemize}[topsep=4pt, itemsep=4pt, partopsep=4pt]
\item Balanced Accuracy ($BA$)~$ = \frac{1}{2}(\frac{TP}{TP + FN} + \frac{TN}{TN+FP})$
\item Precision ($Pr$)~$ = \frac{TP}{TP + FP}$
\item Recall ($Rec$)~$ = \frac{TP}{TP + FN}$
\end{itemize}
Balanced accuracy ($BA$) indicates whether the model identifies the positive and negative classes correctly and penalizes the score in the presence of false positives and negatives.
Precision ($Pr$) indicates the model sensitiveness to false positives.
Finally, recall ($Rec$) indicates whether the model identifies all data points from the positive class.

\subsection{Evaluation}
\label{sec:methodology:eval}

\textbf{RQ1}~\emph{(model performance in a large-scale system)}: We use the \adyen{} dataset (see Table~\ref{tab:subjects:summary}) and apply the process and learning algorithms described in Section~\ref{sec:methodology:ml}.
Unfortunately, the related work either lacks publicly available implementation~\cite{li_where_2020,li_studying_2018} or requires a full re-implementation compatible with our Java code base~\cite{zhu_learning_2015,jia_smartlog_2018}.
For this reason, we use two probabilistic baselines: random guess with $p=50.0\%$ and biased guess with $p=7.7\%$ for the positive class.
The later baseline illustrates a developer that is aware that only $7.7\%$ of the methods are logged in the code base (see Table~\ref{tab:subjects:summary}) and would follow this reasoning for logging decisions.
In addition to balanced accuracy ($BA$), precision ($Pr$), and recall ($Rec$), we also report the confusion matrix for further context.

\textbf{RQ2}~\emph{(the effect of sampling on imbalanced data)}: We add class balancing as a pre-processing step for each learning algorithm on model optimiziation (see Figure~\ref{fig:ml-pipeline}).
For under-sampling, we use Random Under sampling (RUS).
In this setting, class balancing occurs by randomly dropping data points from the majority class, i.e., method should not be logged.
For over-sampling, we use Synthetic Minority Over-sampling TEchnique (SMOTE)~\cite{chawla_smote_2002}.
In this setting, class balancing occurs by generating synthetic data points similar to the points from the minority class.
We compare under/over-sampling with no-sampling ($RQ1$) over the same test split, and report the gain (or loss) on the performance metrics.

\textbf{RQ3}~\emph{(feature importance)}: We inspect the optimized models and rank the most relevant features across models.
The rationale for ranking (rather than reporting the actual values) is that it is unfeasible to directly compare coefficients (for logistic regression) and feature importance (for tree-based models).
For logistic regression, we use the absolute values from the coefficients as a baseline of feature importance.
For the tree-based models, we rank the features with the best mean decreased impurity, i.e., the features that best split the data into the predicted classes.
In addition, since we rely on \emph{after-training} statistics, we also inspect models that used class balancing ($RQ2$).
We report the distribution of the top five most relevant features across models during training.

\textbf{RQ4}~\emph{(to which extent a model trained with open-source data can generalize to an industry setting)}: We use the Apache dataset to train a model and evaluate it against the same test set from the \adyen{} dataset.
We evaluate the performance of the model in two scenarios.
First, we build a training set composed by the dataset of all Apache projects together.
The train set contains \apacheMethods{} data points from the \apacheNum{} projects combined.
Second, we train a model using the dataset of an Apache project individually; therefore, we run the experiment \apacheNum{} times for each project.
We report the performance metrics similarly to $RQ1$ for both scenarios.

\section{Results}
\label{sec:results}

\subsection{\rqA}

\begin{table}[!t]
\caption{Model performance ordered by balanced accuracy ($BA$).}
\centering
\begin{tabular}{@{}p{20mm}@{}rrr|rrrr@{}}
\toprule
\multirow{2}{*}{Model}
    & \multirow{2}{*}{$BA$}
    & \multirow{2}{*}{$Pr$}
    & \multirow{2}{*}{$Rec$}
    & \multicolumn{4}{c}{confusion matrix ($\textrm{N} = 61,906$)} \\

& & & &     $TN$ &    $FP$ &    $FN$ &    $TP$ \\

\midrule
\textbf{Random Forest}       & \textbf{0.79} &  \textbf{0.81} &    \textbf{0.60} & \textbf{56,471} &    \textbf{690} & \textbf{1,892} & \textbf{2,853} \\
Extra Trees         & 0.77 &  0.74 &    0.55 & 56,229 &    932 & 2,145 & 2,600 \\
Decision Tree       & 0.74 &  0.57 &    0.50 & 55,388 &  1,773 & 2,354 & 2,391 \\
AdaBoost            & 0.70 &  0.64 &    0.42 & 56,036 &  1,125 & 2,756 & 1,989 \\
Logistic Reg.       & 0.67 &  0.65 &    0.36 & 56,251 &    910 & 3,050 & 1,695 \\
\midrule
Biased Guess        & 0.50 &  0.08 &    0.08 & 52,764 &  4,397 & 4,363 &   382\\
Random Guess        & 0.50 &  0.08 &    0.50 & 28,595 & 28,566 & 2,386 & 2,359\\

\bottomrule

\end{tabular}
\label{tab:rq1}
\end{table}

Table~\ref{tab:rq1} shows the performance of all evaluated models.
We trained all models using the same training set ($N=247,621$) and evaluate against the same test set ($N=61,906$) from \adyen{}'s dataset.
Both splits preserve the class balancing of the original dataset, i.e., $7.7\%$ (see Table~\ref{tab:subjects:summary}).
On average, the models achieved $73.3 \pm 5.0\%$ of balanced accuracy, $68.1 \pm 9.0\%$ of precision, and $48.6 \pm 9.8\%$ of recall.
The best performing model was Random Forest in all scores.

The confusion matrix provides further context to the results.
For instance, while logistic regression is a simple linear classifier with 65\% of precision, it has the second best rate of $FP$ (the lower, the better).
Note that random guess achieves better scores than guessing based on the label distribution (biased guess); however, the results show that random guess has nearly $\times6.5$ more false positives compared to biased guess.
In practice, that means $\times6.5$ more false alarms to a developer.
Furthermore, while biased guess was useful to optimize $TN$ and $FP$ compared to random guess, it was insufficient to detect true positives.
Nevertheless, all machine learning models achieved better balance of $TN$s and $TP$s.

\begin{mdframed}[hidealllines=true,backgroundcolor=gray!20,innerrightmargin=1mm,innerleftmargin=1mm]
\emph{$RQ1$: The Random Forest model outperforms all other models and probabilistic baselines with 79\% of balanced accuracy, 81\% of precision, and 60\% of recall.}
\end{mdframed}

\subsection{\rqB}

The rationale for class balancing is to better perceive the minority class, i.e., reduction of false negatives in our context, for a better performing model after training.
Table~\ref{tab:rq2} shows the gains and losses of using Synthetic Minority Over-sampling TEchnique (SMOTE) and Random Under Sampling (RUS) compared to training on imbalanced data (as in $RQ1$).

Overall, SMOTE and RUS significantly reduced false negatives which reflect on better balanced accuracy and recall.
However, precision was severely reduced due to the high rate of false positives.
For SMOTE, balanced accuracy and recall increased  by $14.7\pm4.3\%$  and $38.8\pm11.5\%$, respectively while precision reduced by $-27.8\pm5.9\%$.
For RUS, balanced accuracy and recall increased by $16.2\pm3.4\%$ and $43.6\pm6.9\%$, respectively while precision reduced by $-31.1\pm7.3\%$.
The difference between SMOTE and RUS was negligible for logistic regression (less than 1\% in all metrics; hence, the same loss and gain in Table~\ref{tab:rq2}).

Finally, the results show that while sampling strategies improve the perception of the minority class (i.e., increase of recall), the resulting models mislabel data points from the majority class (i.e., decrease of precision).

\begin{table}[!t]
\centering
\caption{Difference of class balancing for model training compared to training on imbalanced data (Table~\ref{tab:rq1}).}
\footnotesize
\begin{tabular}{@{}p{19mm}@{}p{13mm}@{}rrr|rr@{}}
\toprule
Model & Balancing\textsuperscript{a} & $BA$ &  $Pr$ &  $Rec$ & $FP$ & $FN$ \\
\midrule
Random Forest & SMOTE & +0.10 & -0.31 &   +0.26 & +3,538 & -1,227 \\
              & RUS   & +0.12 & -0.40 &   +0.35 & +6,062 & -1,677 \\
Extra Trees   & SMOTE & +0.13 & -0.29 &   +0.34 & +4,313 & -1,624 \\
              & RUS   & +0.15 & -0.36 &   +0.41 & +6,723 & -1,942 \\
Decision Tree & SMOTE & +0.12 & -0.18 &   +0.32 & +4,214 & -1,514 \\
              & RUS   & +0.15 & -0.22 &   +0.40 & +6,046 & -1,912 \\
AdaBoost      & SMOTE & +0.18 & -0.33 &   +0.50 & +8,721 & -2,387 \\
              & RUS   & +0.19 & -0.28 &   +0.49 & +6,722 & -2,343 \\
Logistic Reg. & SMOTE & +0.20 & -0.28 &   +0.52 & +6,028 & -2,444 \\
              & RUS   & +0.21 & -0.28 &   +0.52 & +6,274 & -2,474 \\
\bottomrule
\multicolumn{7}{@{}p{8.3cm}@{}}{
{\textsuperscript{a}RUS = Random under sampling, SMOTE = Synthetic minority over-sampling technique~\cite{chawla_smote_2002}}
}
\end{tabular}
\label{tab:rq2}
\end{table}

\begin{mdframed}[hidealllines=true,backgroundcolor=gray!20,innerrightmargin=1mm,innerleftmargin=1mm]
\emph{$RQ2$: While SMOTE and Random Under Sampling increase balanced accuracy and recall, they penalize precision at a prohibitive increase of false positives.}
\end{mdframed}

\subsection{\rqC}
\label{sec:res:rq3}

Table~\ref{tab:rq3} shows the top five most recurrent features across models.
For all 15 models (five from $RQ1$ and 10 from $RQ2$), we selected the top five features and ranked their frequency across models.
We analyze all models since factors associated to the training process (e.g., scaling, hyper-parameters, and sampling) impact how a model perceives the features.
Column ``Feature'' indicates the feature name, column ``Scope'' indicates whether the metric was computed at method or class level, column ``ranking'' indicates the ranking from the given feature, e.g., ``RFC'' was the third most important features three times (ranking $3\textsuperscript{rd} = 3$), and column ``Total'' sums up the ranking column.
Features are ordered by the column ``Total''.

The feature ``maxNestedBlocks'' was the most recurrent feature ($\times14$), followed by ``SLOC'' ($\times9$) and ``methodsInvokedQty'' ($\times8$).
In the context of object-oriented metrics, ``CBO'' was the most recurring feature ($\times6$), followed by ``WMC'' ($\times5$) and ``RFC'' ($\times3$).
Furthermore, most of the reported metrics relates to some extent to method complexity beyond method length, e.g., depth of method (``maxNestedBlocks''), number of method calls, and coupling between objects (``CBO'').
Finally, it is worth mentioning that feature importance provides a basic framework to debug/analyze what characteristics from the dataset the models learn after training.
It is not meant to be interpreted as a causality analysis.

\begin{table}[!t]
\centering
\caption{Top five most recurrent features across models ordered by ranking and frequency (column ``Total'').}
\begin{tabular}{@{}lc|rrrrr|r@{}}
\toprule
\multirow{2}{*}{Feature}
    & \multirow{2}{*}{Scope}
    & \multicolumn{5}{c|}{ranking}
    & \multirow{2}{*}{Total} \\
               
  & &    {1\textsuperscript{st}} &  {2\textsuperscript{nd}} &  {3\textsuperscript{rd}} & {4\textsuperscript{th}} & {5\textsuperscript{th}} & \\
\midrule

\textbf{maxNestedBlocks} & \textbf{method} &  \textbf{11} &  \textbf{1}   &       &  \textbf{2}   &     &     \textbf{14} \\
               SLOC & method &   1 &  2   &   1   &  3   &  2  &      9 \\
  methodsInvokedQty & method &     &  2   &   1   &  1   &  4  &      8 \\
                CBO & method &   1 &  2   &   2   &  1   &     &      6 \\
     uniqueWordsQty & method &     &  1   &   3   &  2   &     &      6 \\
                WMC & method &     &  3   &   1   &      &  1  &      5 \\
       variablesQty & method &     &  1   &       &  1   &  2  &      4 \\
    type\_interface & class  &   1 &      &   2   &      &     &      3 \\
                RFC & method &     &      &   3   &      &     &      3 \\
  constructor\_True & method &   1 &  1   &       &      &     &      2 \\
        type\_enum  & class  &     &  1   &       &  1   &     &      2 \\
          returnQty & method &     &      &   1   &  1   &     &      2 \\
                CBO & class  &     &      &       &  1   &  1  &      2 \\
constructor\_False  & method &     &      &       &      &  2  &      2 \\       
\bottomrule
\end{tabular}
\label{tab:rq3}
\end{table}

\begin{mdframed}[hidealllines=true,backgroundcolor=gray!20,innerrightmargin=1mm,innerleftmargin=1mm]
\emph{$RQ3$: The depth of method, i.e., the maximum number of nested blocks (``maxNestedBlocks''), ranks as the most relevant feature in 11 out 15 models.}
\end{mdframed}

\subsection{\rqD}
\label{sec:res:rq4}

Table~\ref{tab:rq4} shows the performance of models trained using open-source data and tested on \adyen{}'s test set.
We train the models using Random Forest due to its best performance in previous experiments.
In addition, we explore two scenarios: learning from all Apache projects (upper part) and learning from each Apache project individually (lower part). 

Overall, no open-source data yield better (or similar) scores compared to the original experiment (see Random Forest, Table~\ref{tab:rq1}).
In the first scenario, balanced accuracy, precision, and recall reduced by 18\% ($=0.79-0.61$), 18\% ($=0.81-0.63$), and 36\% ($=0.60-0.24$), respectively.
In the second scenario, ``cloudstack'' achieved better performance; however, balanced accuracy, precision, and recall reduced by 7\% ($=0.79-0.72$), 19\% ($=0.81-0.62$), and 14\% ($=0.60-0.46$), respectively.
Note that ``cloudstack'' yields comparable precision and outperforms the first scenario in terms of recall and balanced accuracy using $86.5\%$ less data.
In contrast to \adyen{}'s training set, ``cloudstack'' has $78.8\%$ less data.

Interestingly, Table~\ref{tab:rq4} also shows five experiments with comparable (and even lower) $FP$s in contrast to \adyen{}'s training set, despite scoring worse.
For instance, all Apache projects together yields $656$ of $FP$s.
The project ``myfaces-tobago'', a user interface library based on JavaServer Faces (JSF), yields $344$ $FP$s and $70\%$ of precision with only $3.8k$ data points.
The remaining three experiments are ``common-beanutils'', a library to dynamically access Java object properties; ``lens'', a data analytics platform; and ``ambari'', a Hadoop management system.
We see that those models were penalized by their low recall due to high $FN$s.
Nevertheless, all models achieved better balance of $TN$s and $TP$s as seen in balanced accuracy compared to random and biased guessing (see Table~\ref{tab:rq1}).

\begin{table}[!t]
\caption{Performance of open-source data as training set on \adyen{}'s test set ordered by balanced accuracy ($BA$).}
\centering
\small
\begin{tabular}{@{}p{25mm}@{}rp{9mm}@{}p{9mm}@{}r|rr@{}}
\toprule
\multicolumn{1}{@{}l}{\multirow{2}{*}{Dataset\textsuperscript{a}}}
    & \multicolumn{1}{@{}c}{Training}
    & \multirow{2}{*}{$BA$}
    & \multirow{2}{*}{$Pr$}
    & \multirow{2}{*}{$Rec$}
    & \multirow{2}{*}{$FP$}
    & \multirow{2}{*}{$FN$}
    \\

& \multicolumn{1}{@{}c}{size} & & &
\\
\midrule
\multicolumn{1}{@{}l}{All Apache data} & \apacheMethods & 0.61 &  0.63 &    0.24 &   656 & 3,606 \\

\midrule
\textbf{cloudstack}  &  \textbf{52,390} & \textbf{0.72} &  \textbf{0.62} & \textbf{0.46} & \textbf{1,348} & \textbf{2,570} \\
zeppelin          &        10,953 & 0.72 &  0.56 &    0.47 & 1,754 & 2,530 \\
oodt              &         6,933 & 0.71 &  0.52 &    0.45 & 1,915 & 2,631 \\
archiva           &         5,995 & 0.69 &  0.51 &    0.41 & 1,831 & 2,822 \\
helix             &         6,787 & 0.68 &  0.46 &    0.40 & 2,233 & 2,870 \\
thrift            &         1,797 & 0.67 &  0.29 &    0.43 & 4,979 & 2,682 \\
sqoop             &         3,080 & 0.67 &  0.51 &    0.38 & 1,716 & 2,961 \\
bookkeeper        &        12,711 & 0.67 &  0.55 &    0.36 & 1,408 & 3,035 \\
nutch             &         3,321 & 0.67 &  0.59 &    0.35 & 1,158 & 3,065 \\
openmeetings      &         4,839 & 0.66 &  0.47 &    0.34 & 1,815 & 3,111 \\
zookeeper         &         5,279 & 0.64 &  0.62 &    0.30 &   890 & 3,319 \\
jmeter            &         8,597 & 0.64 &  0.61 &    0.29 &   875 & 3,357 \\
reef              &         6,150 & 0.63 &  0.48 &    0.29 & 1,454 & 3,388 \\
accumulo          &        25,458 & 0.63 &  0.56 &    0.28 & 1,020 & 3,437 \\
syncope           &        14,915 & 0.63 &  0.52 &    0.27 & 1,174 & 3,455 \\
giraph            &         8,039 & 0.63 &  0.50 &    0.27 & 1,300 & 3,450 \\
storm             &        24,208 & 0.62 &  0.51 &    0.27 & 1,247 & 3,457 \\
tez               &         8,947 & 0.62 &  0.54 &    0.27 & 1,095 & 3,470 \\
knox              &         6,821 & 0.59 &  0.46 &    0.20 & 1,108 & 3,817 \\
myfaces-tobago    &         3,866 & 0.58 &  0.70 &    0.17 &   344 & 3,950 \\
commons-beanutils &         1,176 & 0.58 &  0.66 &    0.16 &   397 & 3,981 \\
lens              &         6,231 & 0.57 &  0.60 &    0.16 &   494 & 3,994 \\
ambari            &        21,997 & 0.56 &  0.51 &    0.13 &   615 & 4,105 \\
\bottomrule
\multicolumn{7}{@{}l}{\textsuperscript{a}Omitted six projects with $Rec<0.10$ for brevity.}
\end{tabular}
\label{tab:rq4}
\end{table}

\begin{mdframed}[hidealllines=true,backgroundcolor=gray!20,innerrightmargin=1mm,innerleftmargin=1mm]
\emph{$RQ4$: The best performing model (based on Apache Cloudstack) yields 72\% of balanced accuracy, 62\% of precision, and 46\% of recall over \adyen{}'s test set. While all Apache data combined yields an underperforming model, it has a low rate of false positives.}
\end{mdframed}

\section{Discussions}
\label{sec:discussion}

\subsection{The Precision-Recall Trade-off}
Our results from $RQ1$ show that it is feasible to leverage code metrics to train a model on the task of log placement at method level.
The best performing model (based on Random Forest) achieved 79\% of balanced accuracy with only 1.1\% of false positives in the test set.
Even the worst performing model (based on Decision Tree) still achieved 2.8\% in false positives.
We observed empirically that all models mislabeled data points from the positive class (i.e., it was expected to be predicted as ``should be logged''); therefore, those models could improve performance by lowering the false negatives rate.
One way to address this issue would be the use of class balancing.
However, as seen in $RQ2$, our experience on using sampling techniques was negative in the sense that recall and balanced accuracy improved at the cost of significant reduction in precision.
We argue that sacrificing precision over recall (or balanced accuracy) is prohibitive in practice: a model with a high rate of false positives could compromise the usefulness of a recommendation tool since it could be a burden for developers to deal with the noisy output.

\subsection{Learning from Code Metrics and Open Source}
We achieved promising results in the context of our industry partner; however, code metrics might not be a good fit for every context since they are closely related to programming style and design patterns.
For instance, in one extreme case, a developer could breakdown a method into several single-line methods. Conversely, several methods could be merged into a larger and complex method.
Those variances in programming styles might undermine the usefulness of code metrics as predictors.
Complementary features might be helpful to capture further nuances and characteristics of logged methods, e.g., the use of a NLP pipeline to learn from the code vocabulary.
This could be also useful in the challenge of reducing false negatives without undermining precision.

Leveraging open-source data can be useful to overcome the ``cold-start'' problem given that the models outperforms random guessing and guessing based on label distribution.
Furthermore, we observed five scenarios where the resulting model achieved a low rate of false positives.
For instance, the model trained with all Apache data had 647 of false positives (1.2\% of the true negative class).
However, deciding which model to choose or how to combine datasets remains a problem.
More research is necessary to understand what factors projects share to help developers to choose an optimal model similarly to past work in transfer-learning for defect prediction~\cite{zimmermann_cross-project_2009}.


\section{Threats to Validity}
\label{sec:ttv}

\textbf{External Validity.}
The main threat to the generalization of our results relates to the scope of our study.
Our exploratory study focused on the code base of our industry partner.
We used open-source data (\apacheNum{} Apache projects) to investigate whether we could leverage training data from public repositories to overcome the ``cold-start'' problem.
It was out of scope to explore the feasibility of our approach on the each open-source project.
In addition, we consider that developers use logging frameworks in the form of method invocations (imperative style).
However, another form of instrumenting source code for logging is through aspect-oriented programming (declarative style).
While declarative logging is not part of development practices in our industry partner, our approach could be extended to support Java annotations for the labeling process (see Section~\ref{sec:methodology:fe-and-labelling}).

\textbf{Internal Validity.}
Our study focuses on logging practices in production-related code, and we classify source files according to their respective paths (see Section~\ref{sec:methodology:subjects}).
While this approach might wrongly classify paths under project-specific settings, we observed that all Apache projects follow standard directory hierarchy of Maven/Gradle projects~\cite{maven_introduction_nodate,gradle_organizing_2020}.
We also use regular expressions to identify log statements.
We encoded patterns based on the in-house logging framework of our industry partner.
To generalize to other the open-source projects, we added patterns similarly to other studies~\cite{kabinna_examining_2018,li_towards_2017,fu_where_2014,chen_characterizing_2017}.
We did not inspect each project individually to fine-tune our identification process.
Another threat to internal validity is the deletion of log-related code on feature extraction to mitigate \emph{data leakage} (see Section~\ref{sec:methodology:fe-and-labelling}).
We implemented a validation step to compare the presence of log statements before and after the removal of log statements to measure the accuracy of our implementation.
In 23 out of 30 projects, we removed 100\% of the log statements identified.
On the other seven projects, the remaining log ratio was lower than 0.5\%.
Based on the numbers, we believe that our log removal process (including guards) meets our expectations without compromising the results.
Finally, our model optimization process consists of 10 trial runs for each model (see Section~\ref{sec:methodology:ml}).
In practice, it would be necessary more trials to cover a representative sample of the search space (e.g., Random Forest has 2,000 possible settings).
However, we observed marginal gains on balanced accuracy for 100 runs in comparison to 10.
For this reason, we keep the number of trials to ``10'' since it achieves an acceptable trade-off of time-cost and improvements over the default hyper-parameters.

\section{Related Work}
\label{sec:related-work}

\subsection{The ``Log Placement'' Problem}
\label{sec:related-work:placement}
Papers addressing the ``log placement'' problem are closely related in purpose to our goal: providing tooling support to help developers to make informed decisions about where to place log statements.
However, they differ in scope, technique, and assumptions.

\textsc{SmartLog}~\cite{jia_smartlog_2018} provides placement recommendations for error log statements by mining patterns of ``log intention'' in source-code using frequent item set mining~\cite{agarwal_fast_1994}.
The underlying assumption is that error log statements are often associated with check conditions involving the return value and arguments of an error-prone function.
However, in the scope of \adyen, the root cause of an invalid state is not limited to an improper parameter or return value of a method. It also includes unexpected interactions between different components from the application stack that might be nontrivial to capture through dependency analysis.
Our process relies only on code metrics extracted from source code.
In a different work, \textsc{LogAdvisor}~\cite{zhu_learning_2015} provides placement recommendations for two types of code blocks: \code{catch} clauses  and \code{if} statements with return value.
However, our analysis in the code base of \adyen{} shows that log statements are placed in many different places rather than only the blocks supported by \textsc{LogAdvisor} (see Section~\ref{sec:motivation:subject-and-logstatements}).

Later work by Li et al.~\cite{li_where_2020} proposes a deep learning architecture based on Long Short-Term Memory (LSTM) units to predict log placement at block level.
In summary, blocks are represented as feature vectors and, based on the sequence of $n$ blocks, the architecture indicates whether the $n+1$ block should be logged or not in a \emph{sequence-to-vector} fashion.
Similarly to our work, they also extract syntactic features based on the occurrence of AST nodes (e.g., number of method calls); however, they also leverage the code vocabulary (referred as semantic features) from variables and method invocations.
Our work mainly differs on the types of models (deep learning architecture versus traditional machine learning) and granularity (block-level versus method-level).
We choose a simpler approach given the low prevalence of log statements on logged methods (see Section~\ref{sec:motivation:subject-and-logstatements}) and the lack of existing tools (or replication packages) publicly available for comparison.

Another study investigates the relationship of the code vocabulary and the presence of log statements~\cite{li_studying_2018}.
The underlying idea is that some system functionalities are more likely to require log statements (e.g., network communication) than others (e.g., getter and setter methods).
The authors use a probabilistic model (latent Dirichlet allocation~\cite{blei_latent_2003}) to extract topics at method level and analyze the correlation between topics and the presence of log statements.
The similarities in context compared to our work encourages further investigation on benefits and costs of using topic modeling in our domain.

\textsc{Log4Perf}~\cite{yao_log4perf_2018} provides placement recommendations for log statements to the specific purpose of improving application performance monitoring in web systems.
It relies on log analysis and source-code instrumentation to build statistical models and to identify code blocks related to ``performance-influencing web requests'', i.e., requests with unstable performance.
Our work differs in context since we do not differentiate the purpose of log statements.

\subsection{Empirical studies in logging engineering}
\label{sec:related-work:others}

Several studies leverage repository mining and issue tracker data to understand how developers conduct logging.
Yuan et al.~\cite{yuan_characterizing_2012} conducted a quantitative study on four C/C++ open-source projects (later replicated by Chen and Jiang~\cite{chen_characterizing_2017} with 21 java projects) showing that developers struggle on implementing and maintaining log-related code.
Li et al.~\cite{li_qualitative_2020} show qualitatively that logging is challenging and developers rely on their own intuition to balance the trade-offs of logging.
Hassani et al.~\cite{hassani_studying_2018} suggest that there is a lack of ownership when dealing with log-related fixes and that maintenance requires the guidance of experts in the code base.
They also show that improper logging has harmful consequences, e.g., excessive volume of log data, missing log statement, and improper log messages.
Similarly to our analysis of logging engineering in the field, we also observe that developers in a industry setting face challenges.

\section{Future Work}

Our results highlight the feasibility of implementing a machine learning-based tool to support \adyen{} developers based on code metrics.
However, there are several operational and usability concerns that must be addressed before making such a tool part of the development workflow.
For instance, usability concerns include what is the most effective way to provide recommendations (e.g., through a plugin in the integrated development environment or through a bot during code review), and how developers react to false positives (e.g., ``What is the acceptable rate of false positives?'' and  ``Would developers notice the difference between models trained with in-house data and models trained with open-source data?'');
operational concerns include understanding what is the best strategy to update models in a fast-paced evolving code base.

All operational and usability concerns are beyond the scope of this work and require long-term use of a mature tool for validation.
However, the positive results encouraged us to engage with our industry partner to design an infrastructure for machine learning-based analysis.
As a first step, we are currently developing a working prototype at \adyen{} (using our use case of log placement) to collect feedback and conduct controlled experiments with developers.
We want to study developers' engagement, what are the best operational and management practices for a recommendation tool, and evaluate different machine learning architectures in an industry setting. 

\section{Conclusion}
\label{sec:conclusion}

In this work, we demonstrate that simple metrics extracted from source code (e.g., depth of a method, coupling between objects, and SLOC) combined with traditional machine learning models are useful to address the important problem of placing log statements in source code.
We measured the performance of different classifiers using different training sets and sampling techniques in a large code base from a global service provider.
In addition, we showed that models based on data from the Apache ecosystem are relevant in the absence of training data in an industry setting given the low prevalence of false positives in different cases.
Our promising results encourage the next steps on studying how developers deal with a machine learning-based recommendation tool integrated into their development workflow on the log placement problem.
All supporting scripts and tools of our study are publicly available to the community.
We believe that this will foster independent auditing, extension, and future collaborations in the area of automated log placement research.

\ifdefined\showIdentityAllowed

\section*{Acknowledgment}
We are grateful to everyone at \adyen{} that contributed directly and indirectly to the accomplishment of this work, specially Andreu Mora for the valuable feedback and insights.
This research is funded by the NWO MIPL project, grant number 628.008.003.

\fi

\bibliographystyle{IEEEtran}
\Urlmuskip=0mu plus 1mu
\bibliography{IEEEabrv,references}

\end{document}